# Ion irradiation and implantation modifications of magneto-ionically induced exchange bias in Gd/NiCoO

Christopher J. Jensen,[a] Alberto Quintana,[a] Mamour Sall,[b,c] Liza Herrera Diez,[b] Junwei Zhang,[d] Xixiang Zhang,[d] Dafiné Ravelosona,[b,c] and Kai Liu[a,*]

[a] *Physics Department, Georgetown University, Washington, DC 20057, USA*
[b] *Centre de Nanosciences et de Nanotechnologies, CNRS, Université Paris-Saclay, Palaiseau 91120, France*
[c] *Spin-Ion Technologies, Palaiseau 91120, France*
[d] *King Abdullah University of Science & Technology, Thuwal 23955-6900, Saudi Arabia*

**Abstract**

Magneto-ionic control of magnetic properties through ionic migration has shown promise in enabling new functionalities in energy-efficient spintronic devices. In this work, we demonstrate the effect of helium ion irradiation and oxygen implantation on magneto-ionically induced exchange bias effect in Gd/Ni$_{0.33}$Co$_{0.67}$O heterostructures. Irradiation using He$^+$ leads to an expansion of the Ni$_{0.33}$Co$_{0.67}$O lattice due to strain relaxation. At low He$^+$ fluence ($\leq 2\times10^{14}$ ions cm$^{-2}$), the redox-induced interfacial magnetic moment initially increases, owing to enhanced oxygen migration. At higher fluence, the exchange bias is suppressed due to reduction of pinned uncompensated interfacial Ni$_{0.33}$Co$_{0.67}$O spins. For oxygen implanted samples, an initial lattice expansion below a dose of $5\times10^{15}$ cm$^{-2}$ is subsequently dominated at higher dose by a lattice contraction and phase segregation into NiO and CoO-rich phases, which in turn alters the exchange bias. These results highlight the possibility of ion irradiation and implantation as an effective means to tailor magneto-ionic effects.

# 1. Introduction

A key application of mesoscopic magnetic materials in information technology[1] is the magneto-electric control of magnetism, aiming to toggle magnetism through pure electric field effects and bypassing the large energy consumption of electric current based magnetic manipulation due to Joule heating.[2-4] Among the different magneto-electric effects, magneto-ionics has shown exciting potentials not only in allowing low-energy magnetic manipulation through ionic migration, but also in enabling a wide variety of magnetic functionalities in a robust and non-volatile fashion.[5-19] For example, magneto-ionics has been successfully used to tuned perpendicular magnetic anisotropy (PMA),[8, 9, 12] exchange bias,[10, 16, 20-22] magnetization,[11, 14, 18, 23-25] superconductivity,[26, 27] spin textures,[28-30] among others. Most of these studies have been based on oxygen ion migration, which is sensitive to interfacial microstructures[31, 32] and strains,[33] and known to be boosted at grain boundaries.[34, 35] Thus precise modification of microstructural properties may strongly influence the magneto-ionic response.

Ion irradiation / implantation is an effective method to modify sample microstructures and in turn their physical properties.[36] For example, He$^+$ irradiation induced strain changes or intermixing effects have been used to suppress perpendicular magnetic anisotropy in Co/Pt and Co/Pd type of systems for magnetic recording media[37, 38] as well as magnetic skyrmions.[39] It has also been used to promote the $L1_0$ ordering in FePt, which has critical applications in heat-assisted magnetic recording.[40] Similarly, paramagnetic-to-ferromagnetic transition is achieved in $Fe_{60}Al_{40}$ alloys due to structural/compositional changes produced by ion irradiation.[41, 42] Additionally, He$^+$ ion irradiation has also been used to tailor magnetic random access memory (MRAM) devices through either intermixing for enhancing domain wall motion[43] or crystallization.[44]

Recently, we have demonstrated a magneto-ionically and electrically tunable exchange bias in Gd/Ni$_{1-x}$Co$_x$O ($x$ = 0.50, 0.67) heterostructures,[20] where neither layer is ferromagnetic (FM) at room temperature. The exchange bias is caused by the spontaneous reduction of interfacial Ni$_{1-x}$Co$_x$O to a thin FM NiCo through the gettering effect of Gd,[10, 23] which then couples to the rest of the antiferromagnetic (AF) NiCoO. The exchange bias is sensitive to the microstructural details of the FM/AF interface, including the layer thicknesses and roughness, stoichiometry, magnetization, etc.[45-47] Previously, ion irradiation / implantation has been shown to be effective in tailoring exchange bias,[48] e.g., oxygen implantation into Co films has been shown to induce nanoscale Co/CoO interfaces at designed depth, leading to enhanced exchange bias.[49] Thus it offers a promising handle to introduce a controlled and tunable degree of microstructural modifications and explore their effects on the magneto-ionically induced exchange bias.

In this work we employ helium ion irradiation and oxygen implantation to tune the magneto-ionic response, and the induced exchange-bias, of Ni$_{0.33}$Co$_{0.67}$O/Gd heterostructures through structural and compositional modifications in the Ni$_{0.33}$Co$_{0.67}$O films. Helium irradiation of varying fluence creates controlled density of inter-atomic displacements, while oxygen implantation induces not only structural changes, but also compositional variations.[50] Exchange bias and magnetic properties of the bilayer are found to vary sensitively on the irradiation / implantation species and fluence.

## 2. Experimental

Thin film samples of Ni$_{0.33}$Co$_{0.67}$O (40 nm) were dc reactively co-sputtered, using elemental Ni and Co targets, onto thermally oxidized SiO$_2$ (285 nm) /Si (100) substrates in an ultrahigh vacuum system with a base pressure < 1.3 ×10$^{-5}$ Pa. Rates were calibrated to achieve a NiO : CoO ratio of 1:2. The oxygen : argon ratio in the sputtering gas was held at 1:15 with a working pressure of 0.33 Pa. A substrate temperature of 500 ºC was used to promote the growth of face centered cubic (fcc) $Ni_{1-x}Co_xO$ and avoid formation of the spinel phase $NiCo_2O_4$.[51]

After the deposition of $Ni_{0.33}Co_{0.67}O$, helium ion irradiation and oxygen implantation were carried out by Spin-Ion Technologies, in particular for $He^+$ using a helium-S system. All implantation/irradiations were performed at room temperature with an ion incidence angle of 7° with respect to the sample surface to avoid any ion channeling. Ion energies (20 keV for $He^+$ and 30 keV for $O_2^+$ that was split into two 15 keV oxygen atoms implanted in the material) were adjusted so that the $He^+$ ions irradiate the NiCoO layer and are implanted deep into the $SiO_2$ substrate while O atoms are implanted in the middle of the $Ni_{0.33}Co_{0.67}O$ layer. These implantation profiles were calculated using the TRIM (Transport of Ions in Matter) program, as part of the SRIM (stopping range of ions in matter) package.[52] We used the Kinchin–Pease damage calculation mode and the default SRIM values of threshold displacement energies for all elements. The employed fluences were 2, 6, 10, 20, 30 and $50 \times 10^{14}$ ions $cm^{-2}$ for helium irradiation and 5, 10, 20 and $50 \times 10^{15}$ $cm^{-2}$ for oxygen atom implantation. The samples were subsequently cleaned using a standard acetone, isopropanol, and deionized water manual rinse for 1 minute in each bath to remove any surface contaminants during transportation, and blow-dried in nitrogen gas. Finally, a 20 nm layer of Gd was dc sputtered onto the $Ni_{0.33}Co_{0.67}O$, followed by a 10 nm Pd capping layer. Each of these layers were grown at room temperature in a 0.67 Pa Ar atmosphere.

Magnetic characterizations were carried out using the Quantum Design MPMS3 Superconducting Quantum Interference Device (SQUID) magnetometer and Princeton Measurements MicroMag3900 vibrating sample magnetometer. Structural characterization was performed by X-ray diffraction (XRD) in a Malvern-Panalytical X'Pert3 MRD system using Cu $K_\alpha$ radiation under θ-2θ and grazing incidence configurations, the former performed using a 1D PixCel line detector with a step size of 0.01° and a total integration time of 4000 s in the 15-60° range. Surface roughness (root mean square) of $He^+$ irradiated $Ni_{0.33}Co_{0.67}O$ samples, prior to Gd deposition, was measured by atomic force microscopy to be within 0.4-0.8nm.

## 3. Results and Discussion

*Helium Irradiation*

Structural characterization of the as-prepared non-irradiated and $He^+$ irradiated $Ni_{0.33}Co_{0.67}O$ samples was carried out using XRD (Fig. 1a). Only the (111) $Ni_{0.33}Co_{0.67}O$ peak is observed on the θ-2θ scans over a 2θ range of 15-60°, indicating a single phase with a textured growth, as no separate NiO or CoO peaks are observed. The (111) reflection is located at 2θ=37.05° for the unirradiated sample, higher than the 36.73° expected for the 1:2 NiO : CoO ratio,[53] indicating that the film is highly strained, and likely caused by the incorporation of interstitial oxygen during the sputtering process.[54] Upon $He^+$ irradiation, as the fluence increases, the (111) peak shifts towards lower 2θ angles, reaching 36.89° for 3× and 5×10$^{15}$ ions cm$^{-2}$, close to that in the unstrained $Ni_{0.33}Co_{0.67}O$. This shows that the cubic lattice has expanded with increasing $He^+$ fluence (Fig. 1b), which is a manifestation of the strain relaxation caused by the $He^+$ irradiation that induces local microstructure modifications.[55, 56] The peak width remains largely constant, except for the largest irradiation fluence of 5 ×10$^{15}$ ions cm$^{-2}$, where a clear peak broadening is observed (Fig. 1b), indicating a reduction of crystallite size.

After the Gd layer is deposited onto the $Ni_{0.33}Co_{0.67}O$, to understand the correlation between structural modifications and their magneto-ionic responses, magnetometry measurements were performed. These as-grown samples all exhibit ferromagnetic hysteresis loops (Fig. 2a), due to the aforementioned interfacial magnetic NiCo layer formed through the spontaneous gettering of Gd.[10, 20] Since Gd ($T_C \sim 292$ K) and $Ni_{0.33}Co_{0.67}O$ are paramagnetic and AF at room temperature, respectively, they do not contribute to the observed ferromagnetism, which is confirmed by magnetometry measurements of the Gd and $Ni_{0.33}Co_{0.67}O$ layers alone (not shown). From these loops, saturation magnetic moment normalized over sample area ($M_S$) is extracted and plotted as function of the irradiation fluence (Fig. 2b). Initially, $M_S$ increases substantially from $1.0 \times 10^{-4}$ emu cm$^{-2}$ in the unirradiated sample to $1.9 \times 10^{-4}$ emu cm$^{-2}$ for the $2 \times 10^{14}$ ions cm$^{-2}$ fluence. This increase can be understood as an increase in ionic migration due to the microstructure modifications favoring the magneto-ionically induced interfacial NiCo. With higher He$^+$ fluence, however, $M_S$ decreases monotonically. As ionic diffusion is enhanced at grain boundaries compared to bulk diffusion,[35] further lattice distortion and increase in defect concentration start to impede oxygen transport,[57] resulting in a hindered transport across the film[58] and less FM NiCo is formed from the redox reaction.

All samples were then heated to 400 K, i.e. above the $Ni_{0.33}Co_{0.67}O$ Néel temperature of 368 K,[59] and cooled to room temperature in a 10 kOe field. Exchange bias is established in all of the samples, as shown by the shifted hysteresis loops (Fig. 2c). In the unirradiated sample, an exchange bias, $H_E$, of -178 Oe is observed with a coercive field, $H_C$, of 448 Oe (black loop). For the sample irradiated at $6 \times 10^{14}$ ions cm$^{-2}$, $H_E$ is decreased slightly to -164 Oe, while $H_C$ dropped precipitously to 187 Oe (red loop). As fluence further increases, $H_C$ continues its abrupt decline while $H_E$ exhibits a slight increase until $1 \times 10^{15}$ ions cm$^{-2}$, beyond which it also starts to decrease quickly (Fig. 2d). This overall trend is reflection of the irradiation-induced modification of the FM/AF interface, where pinned uncompensated AF interfacial spins anchors the exchange bias[60, 61] and interfacial magnetic frustration enhances the coercivity.[62, 63] As discussed above, He$^+$ irradiation impedes the oxygen ion migration and suppresses the FM NiCo formed from the redox reaction, leading to a reduction in $M_S$ as well as $H_C$. In typical exchange bias systems, pinned uncompensated interfacial AF spins that are responsible for exchange bias have been found to be on the order of a few percent of a monolayer,[60] which represents the case for the unirradiated $Ni_{0.33}Co_{0.67}O$/Gd. At low $He^+$ fluence ($\leq 1 \times 10^{15}$ ions cm$^{-2}$), the relatively low density of defects in $Ni_{0.33}Co_{0.67}O$ caused by the irradiation is not sufficient to affect the density of pinned uncompensated interfacial AF spins, thus the exchange bias remains essentially constant, with a slight increase due to the reduction in $M_S$. As the $He^+$ fluence increases to $\geq 2 \times 10^{15}$ ions cm$^{-2}$, more than a monolayer coverage of ions, the higher degree of defects inevitably starts to destroy pinned uncompensated interfacial AF spins, thus lowering the exchange bias.

*Oxygen Implantation*

Grazing incidence XRD measurements of $Ni_{0.33}Co_{0.67}O$ films at various oxygen implantation doses are shown in Fig. 3. After the implantation process, the $Ni_{0.33}Co_{0.67}O$ film preserves the textured growth along the (111) direction for all the evaluated doses. The (111) peak position shows a clear decrease in 2θ value for the lowest dose ($5 \times 10^{15}$ cm$^{-2}$). This decrease in the 2θ peak position is attributed to the generation of structural defects and strain relaxation, as observed for $He^+$ irradiation. With further increase in dose, however, this peak shifts to higher 2θ angles. The incorporation of oxygen into the film at high dose ($\geq 1 \times 10^{16}$ cm$^{-2}$) has strained the film,[54] leading to the observed lattice contraction. This is a counter effect to the oxygen vacancy induced lattice expansion upon oxygen depletion commonly observed in magneto-ionic systems.[23, 26]

Additionally, the peak width noticeably broadens with increasing dose, with a full-width-at-half-maximum (FWHM) of 0.8-1.1°, appreciably wider than the $He^+$ series shown in Fig. 1b.

Upon close inspection, a clear asymmetry is observed at the lower 2θ side of the peaks. This is manifestation of phase segregation into a CoO rich phase with a larger lattice parameter and a NiO rich phase with a smaller lattice parameter, as shown in Fig. 3b for the $5 \times 10^{16}$ cm$^{-2}$ dose sample. The phase segregation is likely caused by the high dose of oxygen implantation as reported in other systems under ion irradiation.[36]

These oxygen implanted samples were first field-cooled from 400 K to room temperature in a 10 kOe field. Interestingly, no appreciable exchange bias is observed at room temperature in these samples, and the coercivity remains relatively constant, ~100 Oe (not shown). The lack of apparent exchange bias can be understood in terms of the CoO and NiO phase segregation: For the non-irradiated sample, the selected FC temperature (400 K) is larger than the nominal Néel temperature, $T_N$, for Ni$_{0.33}$Co$_{0.67}$O (368 K), and thus exchange bias can be established, as shown in Fig. 2; upon oxygen implantation, as the Ni$_{0.33}$Co$_{0.67}$O phase segregates into NiO ($T_N$ =525 K) and CoO ($T_N$ =293 K), this FC procedure fails to induce exchange bias in either case, as the temperature range is too low for NiO and too high for CoO.

To confirm this effect, further experiments were carried out by adjusting the FC temperatures. A new set of samples was measured at room temperature after field cooling in 10 kOe from 540 K, and another set was measured at 270 K after field cooling in 10 kOe from 300 K. In both cases exchange bias is established for most of the samples, as shown in Fig. 4. For the 540 K field cooled case, a significant exchange bias of -115 Oe is observed in the unirradiated sample, which is expected since the FC temperature exceeds the $T_N$ of Ni$_{0.33}$Co$_{0.67}$O. For oxygen implanted samples, $H_E$ decreases substantially to < 30 Oe for dose of $5 \times 10^{15}$ - $2 \times 10^{16}$ cm$^{-2}$, before vanishing at a dose of $5 \times 10^{16}$ cm$^{-2}$. Here the phase separated NiO is primarily responsible for the finite exchange bias. Similar to the He$^+$ case discussed earlier, the high dose oxygen irradiation suppresses the pinned uncompensated interfacial AF spins that are responsible for the exchange bias, leading to much smaller values of $H_E$. For the set of samples field cooled from 300 K to 270 K, the lack of exchange bias in the as-grown, unirradiated film confirms the occurrence of a solid solution and the lack of appreciable CoO spurious phases that may have been produced during the $Ni_{0.33}Co_{0.67}O$ growth. The appearance of finite $H_E$ at oxygen dose of $5\times10^{15}$ - $2\times10^{16}$ cm$^{-2}$ confirms the phase segregation, suggesting that CoO is present in a larger fraction with increasing dose. The fact that $H_E$ values are larger compared to the 540 K series can be attributed to the larger anisotropy constant of CoO, which is responsible for the exchange bias under this FC condition, compared to NiO. Note that a reduction in AF grain size from the implantation could lead to a lowered blocking temperature,[64] similar to the effect of the CoO phase in that field cooling to a lower temperature would be necessary to establish the exchange bias.

## 4. Conclusion

We have demonstrated the suitability of ion irradiation / implantation as an effective tool to modify the magneto-ionic response of $Ni_{0.33}Co_{0.67}O$/Gd heterostructures. For He$^+$ irradiated $Ni_{0.33}Co_{0.67}O$ layer, strain relaxation leads to a lattice expansion. This structural modification initially improves the oxygen diffusion in $Ni_{0.33}Co_{0.67}O$/Gd and enhances the saturation magnetization at low fluence. With increasing fluence the exchange bias is suppressed beyond $1\times10^{15}$ ions cm$^{-2}$, as the density of pinned uncompensated interfacial AF spins starts to decrease. For oxygen implanted samples, initial strain relaxation and lattice expansion, up to a dose of $5\times10^{15}$ cm$^{-2}$, is subsequently overtaken at higher dose by a lattice contraction and phase segregation into NiO and CoO-rich phases, as the incorporation of oxygen also modifies the sample composition. This drastically affects the induced exchange bias effect which resulted from a competition among the higher Néel temperature of NiO, the higher anisotropy of CoO and the implantation-induced structural defects. These results demonstrate the ionic-migration origin of the induced exchange bias, and its sensitivity to structural and compositional modifications through ion irradiation and implantation.

**Acknowledgement**

This work has been supported in part by the NSF (ECCS-1933527), by SMART, one of seven centers of nCORE, a Semiconductor Research Corporation program, sponsored by National Institute of Standards and Technology (NIST), by KAUST (OSR-2019-CRG8-4081), and by a PREMAT CNRS project and a POC In Lab from University of Paris Saclay. The acquisition of a Magnetic Property Measurements System (MPMS3) at GU, which was used in this investigation, was supported by the NSF (DMR-1828420).

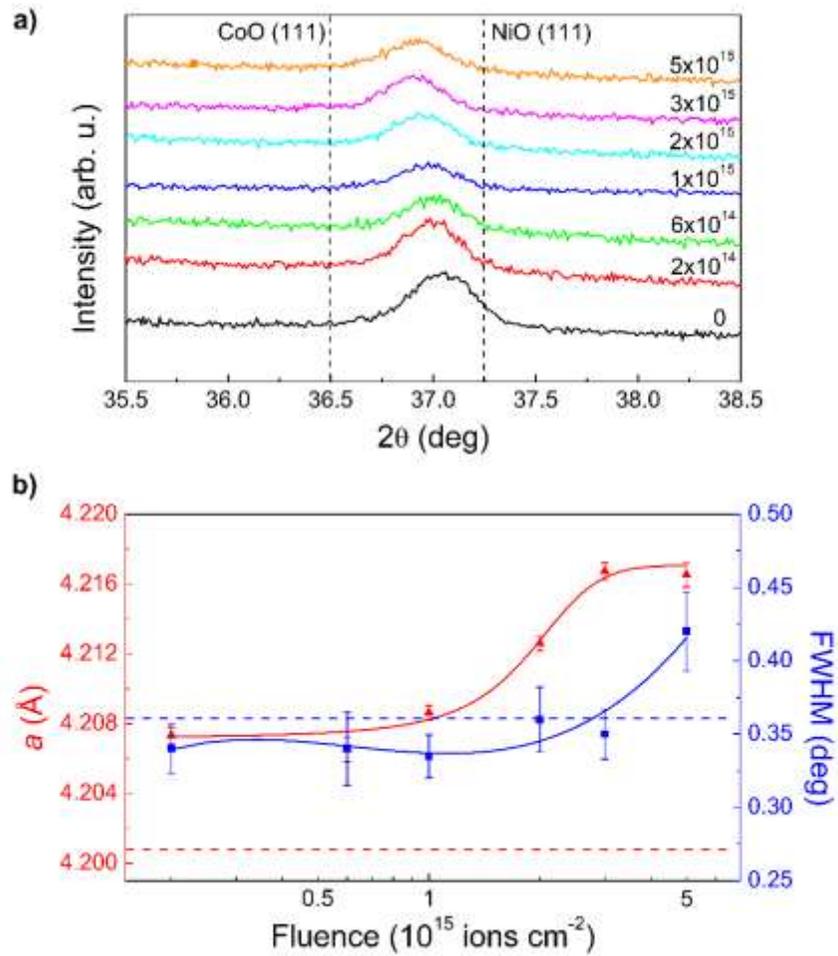

*Fig. 1. a) X-ray diffraction θ-2θ scans of the $Ni_{0.33}Co_{0.67}O$ (111) peak under varying $He^+$ fluence (labeled on the right of each curve in units of ions $cm^{-2}$). b) Dependence of lattice parameter (red) and the full-width-at-half-maximum (FWHM, blue) of the (111) peak in a).*

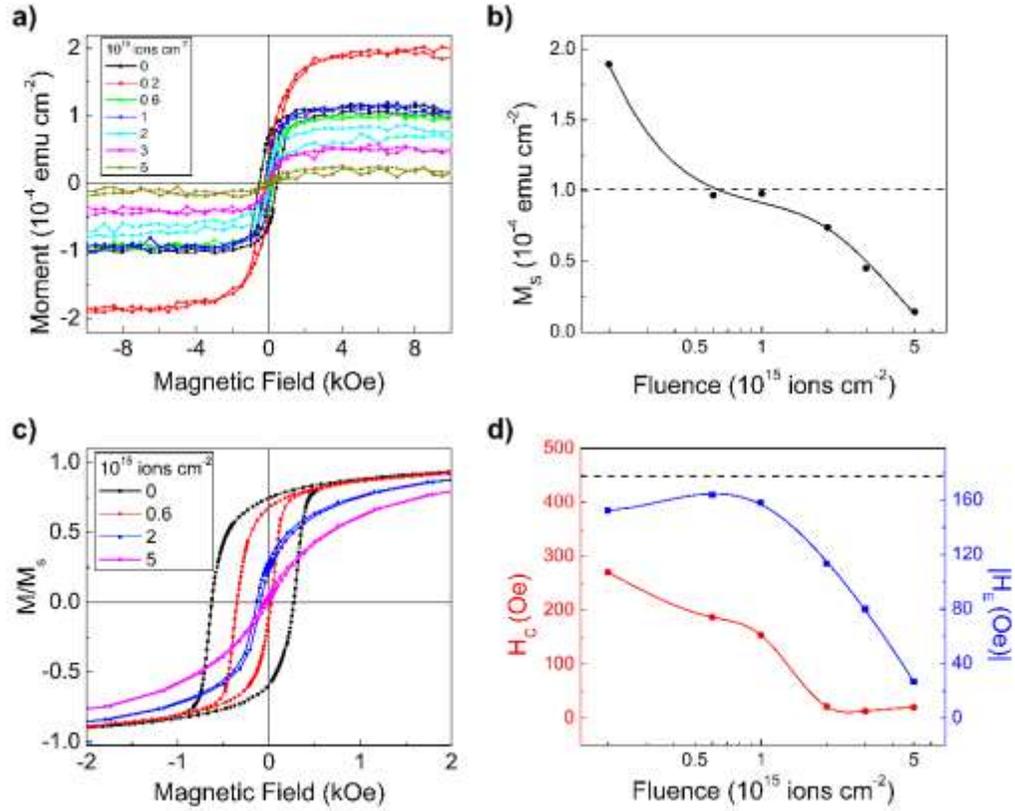

*Fig. 2. a) Room temperature hysteresis loops of as grown $Ni_{0.33}Co_{0.67}O$/Gd films irradiated by varying fluence of $He^+$. b) Saturation moment normalized by sample area as a function of $He^+$ fluence. c) Hysteresis loops of $Ni_{0.33}Co_{0.67}O$/Gd films irradiated by varying fluence of $He^+$ after the field cooling process. d) Coercive field (red) and exchange bias (blue) after the field cooling process as a function of $He^+$ fluence. Dashed lines in b) and d) correspond to values for unirradiated $Ni_{0.33}Co_{0.67}O$/Gd film.*

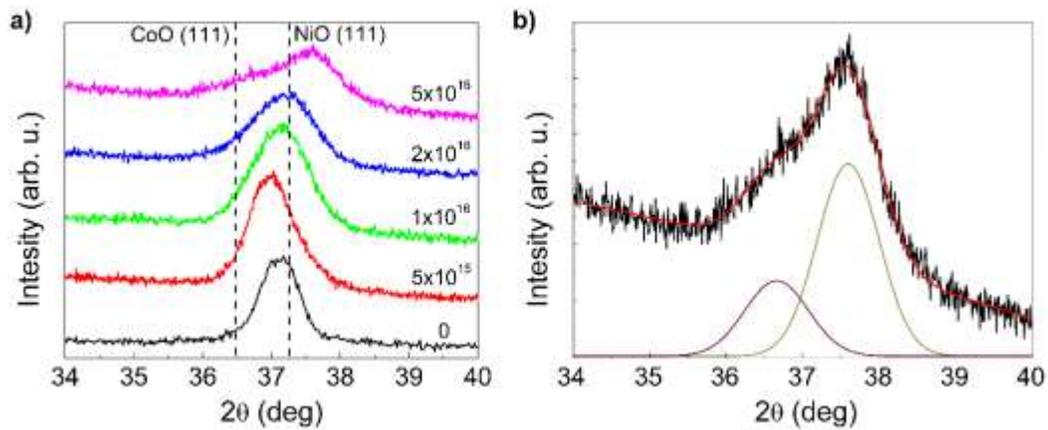

*Fig. 3. a) X-ray diffraction grazing incidence scans of the $Ni_{0.33}Co_{0.67}O$ (111) peak under varying oxygen dose (labeled on the right of each curve in units of atoms/ $cm^{-2}$). b) Deconvolution of the (111) peak for the $5\times10^{16}$ $cm^{-2}$ dose of implanted sample measured using grazing incidence X-ray diffraction. The raw data is shown in black, fit data in red, and the two deconvoluted peaks in black and green, corresponding to CoO and NiO, respectively.*

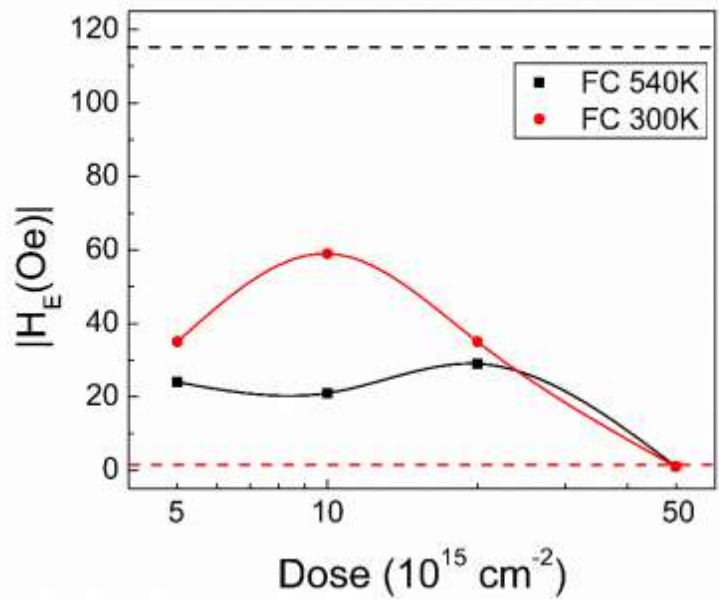

*Fig. 4. Exchange field of oxygen implantation samples vs dose, measured at RT after field cooling from 540 (black), and measured at 270 K after field cooling from 300K (red). Dashed lines indicate corresponding exchange bias values for the unirradiated samples.*